\documentstyle[preprint,eqsecnum,epsfig,aps]{revtex} % from Nikos
\begin{document}
%\special{header=draft.ps}
%
\def\theequation{\arabic{equation}}%
\def\ppbar{$p\overline{p} $}            %ppbar
\newcommand{\rstev}{\mbox{$\rs = \T{1.8}$}}
\newcommand{\T}[1]{\mbox{#1 TeV}}
\newcommand{\rs}{\mbox{$\sqrt{s}$}}
\boldmath
\title{The $b\overline{b}$ Production Cross Section
and Angular Correlations\\
in \ppbar\ Collisions at \rstev}
\unboldmath

% took next line out for draft
%\input{list_of_authors.tex}
% LIST_OF_AUTHORS.TEX                 4/12/00            
%
\author{                                                                      
%% names begin here                                                           
B.~Abbott,$^{48}$                                                             
M.~Abolins,$^{45}$                                                            
V.~Abramov,$^{21}$                                                            
B.S.~Acharya,$^{15}$                                                          
D.L.~Adams,$^{55}$                                                            
M.~Adams,$^{32}$                                                              
V.~Akimov,$^{19}$                                                             
G.A.~Alves,$^{2}$                                                             
N.~Amos,$^{44}$                                                               
E.W.~Anderson,$^{37}$
% ...
\\                                                                            
\vskip 0.30cm                                                                 
\centerline{(D\O\ Collaboration)}                                             
\vskip 0.30cm                                                                 
}                                                                             
\address{                                                                     
\centerline{$^{1}$Universidad de Buenos Aires, Buenos Aires, Argentina}       
\centerline{$^{2}$LAFEX, Centro Brasileiro de Pesquisas F{\'\i}sicas,         
                  Rio de Janeiro, Brazil}                                     
\centerline{$^{3}$Universidade do Estado do Rio de Janeiro,                   
                  Rio de Janeiro, Brazil}                                     
\centerline{$^{4}$Institute of High Energy Physics, Beijing,                  
                  People's Republic of China}                                 
\centerline{$^{5}$Universidad de los Andes, Bogot\'{a}, Colombia}             
\centerline{$^{6}$Charles University, Prague, Czech Republic}                 
\centerline{$^{7}$Institute of Physics, Academy of Sciences, Prague,          
                  Czech Republic}                                             
\centerline{$^{8}$Universidad San Francisco de Quito, Quito, Ecuador}         
\centerline{$^{9}$Institut des Sciences Nucl\'eaires, IN2P3-CNRS,             
                  Universite de Grenoble 1, Grenoble, France}                 
\centerline{$^{10}$CPPM, IN2P3-CNRS, Universit\'e de la M\'editerran\'ee,     
                  Marseille, France}                                          
\centerline{$^{11}$LPNHE, Universit\'es Paris VI and VII, IN2P3-CNRS,         
                  Paris, France}                                              
\centerline{$^{12}$DAPNIA/Service de Physique des Particules, CEA, Saclay,    
                  France}                                                     
\centerline{$^{13}$Panjab University, Chandigarh, India}                      
\centerline{$^{14}$Delhi University, Delhi, India}                            
\centerline{$^{15}$Tata Institute of Fundamental Research, Mumbai, India}     
\centerline{$^{16}$Seoul National University, Seoul, Korea}                   
\centerline{$^{17}$CINVESTAV, Mexico City, Mexico}                            
\centerline{$^{18}$Institute of Nuclear Physics, Krak\'ow, Poland}            
\centerline{$^{19}$Institute for Theoretical and Experimental Physics,        
                   Moscow, Russia}                                            
\centerline{$^{20}$Moscow State University, Moscow, Russia}                   
\centerline{$^{21}$Institute for High Energy Physics, Protvino, Russia}       
\centerline{$^{22}$Lancaster University, Lancaster, United Kingdom}           
\centerline{$^{23}$University of Arizona, Tucson, Arizona 85721}              
\centerline{$^{24}$Lawrence Berkeley National Laboratory and University of    
                  California, Berkeley, California 94720}                     
\centerline{$^{25}$University of California, Davis, California 95616}         
\centerline{$^{26}$California State University, Fresno, California 93740}     
\centerline{$^{27}$University of California, Irvine, California 92697}        
\centerline{$^{28}$University of California, Riverside, California 92521}     
\centerline{$^{29}$Florida State University, Tallahassee, Florida 32306}      
\centerline{$^{30}$University of Hawaii, Honolulu, Hawaii 96822}              
\centerline{$^{31}$Fermi National Accelerator Laboratory, Batavia,            
                   Illinois 60510}                                            
\centerline{$^{32}$University of Illinois at Chicago, Chicago,                
                   Illinois 60607}                                            
\centerline{$^{33}$Northern Illinois University, DeKalb, Illinois 60115}      
\centerline{$^{34}$Northwestern University, Evanston, Illinois 60208}         
\centerline{$^{35}$Indiana University, Bloomington, Indiana 47405}            
\centerline{$^{36}$University of Notre Dame, Notre Dame, Indiana 46556}       
\centerline{$^{37}$Iowa State University, Ames, Iowa 50011}                   
\centerline{$^{38}$University of Kansas, Lawrence, Kansas 66045}              
\centerline{$^{39}$Kansas State University, Manhattan, Kansas 66506}          
\centerline{$^{40}$Louisiana Tech University, Ruston, Louisiana 71272}        
\centerline{$^{41}$University of Maryland, College Park, Maryland 20742}      
\centerline{$^{42}$Boston University, Boston, Massachusetts 02215}            
\centerline{$^{43}$Northeastern University, Boston, Massachusetts 02115}      
\centerline{$^{44}$University of Michigan, Ann Arbor, Michigan 48109}         
\centerline{$^{45}$Michigan State University, East Lansing, Michigan 48824}   
\centerline{$^{46}$University of Nebraska, Lincoln, Nebraska 68588}           
\centerline{$^{47}$Columbia University, New York, New York 10027}             
\centerline{$^{48}$New York University, New York, New York 10003}             
\centerline{$^{49}$University of Rochester, Rochester, New York 14627}        
\centerline{$^{50}$State University of New York, Stony Brook,                 
                   New York 11794}                                            
\centerline{$^{51}$Brookhaven National Laboratory, Upton, New York 11973}     
\centerline{$^{52}$Langston University, Langston, Oklahoma 73050}             
\centerline{$^{53}$University of Oklahoma, Norman, Oklahoma 73019}            
\centerline{$^{54}$Brown University, Providence, Rhode Island 02912}          
\centerline{$^{55}$University of Texas, Arlington, Texas 76019}               
\centerline{$^{56}$Texas A\&M University, College Station, Texas 77843}       
\centerline{$^{57}$Rice University, Houston, Texas 77005}                     
\centerline{$^{58}$University of Washington, Seattle, Washington 98195}       
}                                                                             
%end                                                                          
%\date{\today}
%% end list of authors

\maketitle

\begin{abstract}
We present measurements of the $b\overline{b}$ production cross section
and angular correlations using
the D\O\ detector at the Fermilab Tevatron $p\overline{p}$ Collider operating
at $\sqrt{s}$ = 1.8 TeV. The $b$ quark production cross section for
$|y^b|~<~1.0$ and $p_T^b~>~6$ GeV/$c$ is extracted
from single muon and dimuon data samples.
The results agree in shape with the next-to-leading order QCD 
calculation of heavy flavor production but are greater than the central values of 
these predictions.
The angular correlations between $b$ and $\overline{b}$ quarks, measured
from the azimuthal opening angle between their decay muons, also agree
in shape with the next-to-leading order QCD prediction.  \\


\end{abstract}

%\vspace{0.5cm}
%\vspace{0.3cm}
%%\vspace{0.1cm}
%%\hspace{1.57cm}PACS numbers: 13.65.Fy, 12.38.Qk, 13.85.Ni, 13.85.Qk -- 14.65.Hq, 13.85.Qk 
 
%\vspace{0.5cm}
%\vspace{0.3cm}
%%\vspace{0.1cm}
%%\begin{center}
%%Draft Version Final 3.0 - Post Collaboration Review
%%\end{center}

%%\twocolumn

\newpage

%\section*{Introduction}

Measurements of the $b$ quark production cross section and $b\overline{b}$
correlations in $p\overline{p}$ collisions provide an important test of
perturbative quantum chromodynamics (QCD) at next-to-leading order (NLO). The measured $b$ quark
production cross section at 
$\sqrt{s}$ = 1.8 TeV\cite{cdfb1,d0b1,d0b2,cdfb5} is systematically larger than the central values
of the NLO QCD predictions\cite{nde,mnr}.
%but data and theory are in agreement within
%experimental and theoretical uncertainties.

Measurements of $b\overline{b}$ correlations such as the
azimuthal opening angle between $b$ and $\overline{b}$ quarks allow 
additional details of $b$ quark production to be tested since these quantities 
are sensitive to the relative contributions of different production
mechanisms to the total cross section.
Two measurements of $b\overline{b}$ angular correlations using the azimuthal opening
angle of muons from the heavy quark decays,
one at $\sqrt{s}$ = 1.8 TeV\cite{cdfb4} and another at
$\sqrt{s}$ = 630 GeV\cite{ua1b1}, are in qualitative agreement with the NLO QCD predictions.
A different measurement at $\sqrt{s}$ = 1.8 TeV using the azimuthal opening
angle between a muon from $B$ meson decay and the $\overline{b}$ jet shows\cite{cdfb5}
qualitative differences with
the predictions, while a direct measurement\cite{cdfrap} of $b\overline{b}$ rapidity correlations
is found to be in agreement with the NLO QCD predictions.

In this paper we provide an additional measurement of the
$b$ quark production cross section and $b\overline{b}$ angular correlations.
The analysis makes use of the fact that the semileptonic decay
of a $b$ quark results in a lepton (here a muon) associated with a jet.
We use a sample of dimuons and their associated jets to tag 
both $b$ and $\overline{b}$ quarks. By tagging both the $b$ and $\overline{b}$ quarks,
we are able to significantly reduce the number of background events in our data sample.
Also included is a revised measurement of the $b$ quark production
cross section based on an earlier D\O\ analysis\cite{d0b1} using the inclusive single
muon measurement. 

%\section*{Detector and Event Selection}

The D\O\ detector and trigger system are described in detail elsewhere\cite{d0}.
The central muon system consists of three layers of proportional drift tubes and
a magnetized iron toroid located between the first two layers. The muon detectors
provide a measurement of the muon momentum with a resolution parameterized by
$\delta(1/p)/(1/p)= 0.18(p-2)/p\oplus 0.008p$, with $p$ in GeV/$c$. The
calorimeter is used to measure both the minimum ionizing energy associated with
the muon track and the electromagnetic and hadronic activity associated with
heavy quark decay. The total thickness of the calorimeter plus toroid in the
central region varies from 13 to 15 interaction lengths, which reduces the
hadronic punchthrough in the muon system to less than 0.5\% of
low transverse momentum muons from all sources. The energy resolution for jets is approximately
80\%/$\sqrt{E({\rm GeV})}$.

The data used in this analysis were taken during the 1992--1993 run of the
Fermilab Tevatron collider and correspond to a total integrated luminosity $\int
{\cal L}dt=6.5\pm 0.4$ pb$^{-1}$. The dimuon data were collected using  a
multilevel trigger requiring at least one reconstructed muon with transverse
momentum $p_T^{\mu}>3$ GeV/$c$ and at least one reconstructed jet with
transverse energy $E_T>10$ GeV.

The events are then fully reconstructed offline
and subjected to event selection criteria. The offline analysis requires two
muons with $p_T^{\mu}>4$ GeV/$c$ and pseudorapidity $|\eta^{\mu}|<0.8$. In
addition, both muon tracks have to be consistent with originating from the reconstructed event
vertex and deposit $>1$ GeV of energy in the calorimeter. Each muon is also
required to have an associated jet with $E_T>12$ GeV within a cone of radius ${\cal R}
= \sqrt{(\Delta\eta)^2+(\Delta\phi)^2} < 0.8$. The jet energies are measured using a cone of
${\cal R}=0.7$. Finally, muon candidates in the region
$80^{\circ}<\phi^{\mu}<110^{\circ}$ are excluded due to poor chamber
efficiencies near the Main Ring beam pipe.

Further selection criteria are placed on the dimuon candidates to reduce backgrounds to
$b\overline{b}$ production. The invariant mass of the dimuons is restricted to
the range $6<m^{\mu\mu}<35$ GeV/$c^2$. The lower limit removes dimuons resulting
from the cascade decay of single $b$ quarks and from $J/\psi$ resonance decays, while the
upper limit reduces the number of dimuons due to $Z$ boson decays. An opening
space angle requirement of
%$\Delta\phi_{3D}<165^{\circ}$
$<165^{\circ}$ between the muons is also applied to remove
contamination from cosmic ray muons. A total of 397 events pass all selection
criteria.

The trigger and offline reconstruction efficiencies are determined from Monte
Carlo event samples. Events generated with the {\footnotesize\rm ISAJET\normalsize}\cite{isajet}
Monte Carlo are passed
through a {\footnotesize\rm GEANT\normalsize}\cite{geant} simulation of
the D\O\ detector followed by trigger simulation and reconstruction
programs. Trigger and some offline efficiencies found in this way are
crosschecked by using appropriate data samples. The overall acceptance times
efficiency as a function of the higher (leading) muon $p_T$ in the event increases from about
1\% at 4 GeV/$c$ to a plateau of 9\% above 15 GeV/$c$.
We define the leading muon in the event as the muon with the greater value of $p_T^{\mu}$.

%\section*{Signal and Background Determination}

In addition to $b\overline{b}$ production, dimuon events in the invariant mass range of 6--35
GeV/$c^2$ can also arise from other sources. These processes include
semileptonic decays of $c\overline{c}$ pairs, events in which one or both of
the muons are produced by in-flight decays of $\pi$ or $K$ mesons, Drell-Yan
production, and $\Upsilon$ resonance decays. Muons from the Drell-Yan process and
$\Upsilon$ decays are not expected to have jets associated with them. Monte
Carlo estimates normalized to the measured Drell-Yan and $\Upsilon$ cross
sections\cite{albthesis} show that less than one event is expected to contribute
to the final data sample from these two sources. An additional source of dimuon events
is cosmic ray muons passing through the detector.

To extract the $b\overline{b}$ signal, we use a maximum likelihood fit with
four different input distributions. The input distributions are chosen based on
their effectiveness in distinguishing between the different sources of dimuon
events. We use the transverse momenta of the leading and trailing muons
relative to their associated jet axes ($p_T^{\rm rel}$), the fraction of
longitudinal momentum of the jet carried by the leading muon divided by the jet
$E_T$ ($r_z$), and the reconstructed time of passage ($t_0$) of the leading muon
track through the muon chambers with respect to the beam crossing time. The
variable $t_0$ is used to identify the cosmic ray muon background, which is
not expected to be in time with the beam crossing. Monte Carlo studies show
that the variables $p_T^{\rm rel}$ and $r_z$ help to discriminate between background
and $b\overline{b}$ production. For both variables, the jet energy is defined to
be the vector sum of the muon energy and the jet energy measured in the calorimeter
less the expected minimum ionizing energy of the muon deposited in the calorimeter.

The $p_T^{\rm rel}$ and $r_z$ distributions for $b\overline{b}$, $c\overline{c}$, and
$b$ or $c$ plus $\pi/K$ decay are modeled using the {\footnotesize\rm
ISAJET\normalsize} Monte Carlo. Each of these samples is processed with a
complete detector, trigger, and offline simulation. The distributions for
$b$ quark decays includes both direct ($b\rightarrow\mu$) and sequential
($b\rightarrow c\rightarrow \mu$) decays. The distributions for $c\overline{c}$
and a $c$ quark plus a $\pi$ or $K$ decay are very similar, so both
contributions are fit to the same function. The distributions for $t_0$ are
obtained from two different sources. The $t_0$ distribution for cosmic ray
muons is obtained from data collected between collider runs using cosmic ray triggers.
For beam-produced muons, $t_0$ is measured using muons from $J/\psi$ decays.

Figure~1 shows the result of the maximum likelihood fit for
$p_T^{\rm rel}$ of the leading muon and $r_z$. Included in Fig.~1 are the
contributions from each of the major sources of dimuon events.  The $b\overline{b}$
contribution to the final data sample is found to be 45.3$\pm$5.8\%. The other
fractions fit to the data set consist of $b$ quark plus $\pi/K$ decay
%(\mbox{$37.9^{+5.5}_{-5.7}$}\%), $c\overline{c}$ production
%(\mbox{$14.0^{+3.9}_{-3.7}$}\%), and cosmic ray muons
%(\mbox{$2.8^{+1.7}_{-1.5}$}\%). From the fit, we obtain the number of
(37.9$\pm$5.6\%), $c\overline{c}$ production
(14.0$\pm$3.8\%), and cosmic ray muons
(2.8$\pm$1.6\%). From the fit, we obtain the number of
$b\overline{b}$ events per bin as a function of $p_T^{\mu}$ of the leading
muon and as a function of the difference in azimuthal angle between the two
muons, $\Delta\phi^{\mu\mu}$.

The systematic errors on the number of
$b\overline{b}$ events per bin (8\%) are estimated by varying the input
distributions to the maximum likelihood fit within reasonable bounds. 
As a crosscheck of the fitting procedure, we calculate the fraction of
events originating from $b\overline{b}$ production using appropriately
normalized Monte Carlo samples. Good agreement is found between
the Monte Carlo calculated fraction and that found from the maximum likelihood fit to
the data. The fractions agree as a function of 
both $p_T$ of the leading muon and $\Delta\phi^{\mu\mu}$. 
A complete description of the fitting procedure can be found in
Ref.~\cite{dfthesis}.

%\section*{Results and Discussion}

The dimuon cross section originating from $b\overline{b}$ production is
calculated using \begin{equation}
\frac{d\sigma_{b\overline{b}}^{\mu\mu}}{dx} = \frac{1}{\Delta
x} \frac{N_{b\overline{b}}^{\mu\mu}(x)f_p(x)}{\epsilon(x)\int {\cal L}dt}, \label{eqn:dimu} \end{equation}
where $x$ is either the $p_T$ of the leading muon or $\Delta\phi^{\mu\mu}$.
Here, $\epsilon$ is the total efficiency, $\int {\cal L}dt$ is the
integrated luminosity, $N_{b\overline{b}}^{\mu\mu}$ is the number of $b\overline{b}$ events determined from
the fit, and $f_p$ is an unfolding factor to account for
smearing caused by the muon momentum resolution. An unfolding technique\cite{bayes}
is used to determine $f_p$. The factor $f_p$ varies from 0.78 at 
low $p_T^{\mu_1}$ to 0.93 in the 
highest $p_T^{\mu_1}$ bin ($\mu_1$ is the leading 
muon in the event) and takes into account our invariant mass and $p_T^{\mu}$
requirements.
The systematic uncertainty associated with $f_p$ is found to be $p_T^{\mu}$ dependent
and varies from 13\% to 22\%. 

Figure~2(a) shows the result of the cross section calculation as a function of
$p_T^{\mu_1}$  for
$4<p_T^{\mu}<25$ GeV/$c$, $|\eta^{\mu}|<0.8$, and $6<m^{\mu\mu}<35$ GeV/$c^2$.
The total systematic error is found to be $p_T^{\mu_1}$ dependent, ranging from
25\% to 31\%. This includes uncertainties from the trigger efficiency (19\%), 
offline selection efficiency (5\%), maximum likelihood fit (8\%), 
momentum unfolding (13--22\%), and integrated luminosity (5\%).

The theoretical curve of Fig.~2(a) is determined using
the {\footnotesize\rm HVQJET\normalsize}\cite{hvqjet} Monte Carlo.
{\footnotesize\rm HVQJET\normalsize} is an implementation of the NLO calculation of Ref.~\cite{mnr} (MNR) for
$b\overline{b}$ production.  It uses the MNR parton level generator and a
modified version of {\footnotesize\rm ISAJET\normalsize} for hadronization,
particle decays, and modeling of the underlying event.
The particle decays are based on the {\footnotesize\rm ISAJET\normalsize} implementation\cite{isajet2} of
the CLEO decay tables.
In {\footnotesize\rm HVQJET\normalsize} the MNR prediction is realized by combining
parton level events having negative weights  with
those having positive weights and similar topologies.  The prediction shown
is the NLO calculation and includes all four $gg$, $gq$, $g\overline{q}$, and $q\overline{q}$ initiated
subprocesses with
$m_b({\rm pole~mass})=4.75$ GeV/$c^2$. The MRSR2\cite{mrsr2} parton distribution 
functions (PDFs) are used with $\Lambda_5$ = 237 MeV.

The shaded region in Fig.~2(a) shows the combined systematic and statistical
error from the {\footnotesize\rm HVQJET\normalsize} prediction ($^{+74}_{-50}$\%). This error is dominated
by the uncertainty associated with the MNR prediction and is
determined by varying the mass of the $b$ quark between 4.5 GeV/$c^2$ and 5.0
GeV/$c^2$, and the factorization and renormalization scales, taken to be equal, between
$\mu_0/2$ and $2\mu_0$, where $\mu_0^2=m_b^2+\langle p_T^b\rangle^2$.
Additional systematic errors include those associated with the PDFs (20\%),
the Peterson fragmentation function\cite{peterson} (8\%), the $B$
meson semileptonic branching fraction (7\%), and the muon decay spectrum from $B$
mesons (20\%). Varying these parameters does not appreciably change the 
shape of the prediction.
The Monte Carlo statistical errors are less than 10\%.  

To extract the $b$ quark cross section from the dimuon data, we employ a method
first used by UA1\cite{ua1b2} and subsequently used by CDF\cite{cdfb1} and
D\O\ \cite{d0b1}. Since a correlation exists between the $p_T$ of the muon
produced in a $b$ quark decay and the parent $b$ quark $p_T$, cuts applied
to the muon $p_T$ in the data are effectively $b$ quark $p_T$ cuts. For a 
set of kinematic cuts, which include cuts on the transverse momentum of the 
muons, we define
$p_T^{\rm min}$ as that value of the $b$ quark $p_T$ where 90\% of the accepted
events have $b$ quark transverse momentum greater than $p_T^{\rm min}$. The
$b$ quark cross section is then calculated as
\begin{equation}
\sigma_b(p_T^b>p_T^{\rm min}) = \sigma_{b\overline{b}}^{\mu\mu}(p_T^{\mu_1})
\frac{\sigma_b^{\rm MC}}{\sigma_{b\overline{b}\rightarrow\mu\mu}^{\rm MC}},
\end{equation}
where $\sigma_{b\overline{b}}^{\mu\mu}(p_T^{\mu_1})$ is the
measured dimuon cross section of Eq.~(\ref{eqn:dimu}) integrated over different
intervals of $p_T^{\mu_1}$, $\sigma_b^{\rm MC}$ is the total Monte Carlo $b$ quark
cross section for $p_T^b>p_T^{\rm min}$ (where $|y^b|<1.0$ and no cut on $y^{\overline{b}}$),
and $\sigma_{b\overline{b}\rightarrow\mu\mu}^{\rm MC}$ is the Monte Carlo cross section
for dimuon production with the same requirements used to select the data set.
For each interval of $p_T^{\mu_1}$, $p_T^{\rm min}$ and $\sigma^{\rm MC}$ are calculated using
{\footnotesize\rm HVQJET\normalsize}. Combining the uncertainties of the measured dimuon cross section 
with those associated with extracting the $b$ quark cross section, we obtain
a total systematic uncertainty of 34--38\% on the measured $b$ quark cross section.  
The latter uncertainties are associated with $b$ quark fragmentation:
Peterson fragmentation function; semileptonic 
branching fraction; and muon decay spectrum with the magnitudes 
noted above.

Figure~2(b) shows the $b$ quark production cross section for the
rapidity range $|y^b|<1.0$ as a function of $p_T^{\rm min}$.
The NLO QCD prediction is computed using Ref. \cite{mnr}   
with $m_b({\rm pole~mass})=4.75$ GeV/$c^2$ and the MRSR2 PDFs.
The theoretical uncertainty of $^{+47}_{-28}$\% results from varying
the mass of the $b$ quark and the factorization and renormalization 
scales as described above and is dominated by the variation of the scales.
The ratio of the data to the central NLO QCD prediction
is approximately three over the entire $p_T^{\rm min}$ range covered.

Also shown in Fig.~2(b) is a revised result based on the
previous inclusive single muon measurement from D\O\ \cite{d0b1}.
In light of revised $B$ meson decay modes and 
Monte Carlo improvements, the cross section is re-evaluated by using {\footnotesize\rm HVQJET\normalsize}
to calculate new values of $\sigma_b^{\rm MC}/\sigma_{b\rightarrow\mu}^{\rm MC}$
for extraction of the $b$ quark cross section from the
measured inclusive single muon spectrum. 
In addition, the high $p_T$ inclusive muon data ($p_T^{\mu}>$12 GeV/$c$) are excluded due to
large uncertainties in the cosmic ray muon background subtractions. 
The resulting increase in the $b$ quark cross section
is primarily caused by the new $B$ meson decay modes and lower semileptonic branching fractions\cite{isajet2}.
The re-evaluated cross section supersedes that of Ref.~\cite{d0b1}. The tabulated data for the dimuon and inclusive
single muon data sets can be found in Tables~I and~II.

%The results of the $b$-quark production cross section as a function of
%$p_T^{min}$, for $|y^b|<1.0$, are shown in Fig.~\ref{fig:sigma}(b). Also
%shown is a previous measurement of the $b$-quark cross section from D\O\
%\cite{d0b1}. The $b$-quark cross section result from Ref.~\cite{d0b1} was re-evaluated
%using the most recent hadronization and decay model of~\cite{hvqjet}.
%Using this Monte Carlo, we calculated new values to extract the 
%$b$-quark cross section from the inclusive
%muon spectrum. In addition, the high $p_T^{\mu}$ data points are withheld due to
%uncertainties in the background subtraction. This data supercedes that of Ref.~\cite{d0b1}.
  
%\vspace{-0.3cm}
The differential $b\overline{b}$ cross section,
$d\sigma_{b\overline{b}}^{\mu\mu}/d\Delta\phi^{\mu\mu}$, gives further
information on the underlying QCD production mechanisms. The azimuthal opening
angle between $b$ and $\overline{b}$ quarks (or between their decay muons) is sensitive
to the contributing production mechanisms. These contributions are the
leading order (LO) subprocess, flavor creation, and the next-to-leading order 
subprocesses, gluon splitting and flavor excitation. There are also 
contributions from interference terms.

%The $\Delta\phi^{\mu\mu}$ spectrum calculated in Eq.~(\ref{eqn:dimudphi}) is first acceptance corrected
%for the three-dimensional opening angle cut.
%The $\Delta\phi^{\mu\mu}$ spectrum is then fit with each of the heavy flavor contributions, separately
%and without constraint. The results of this fit are shown in Fig.~\ref{fig:lonlofit}. Good agreement
%is found in the amount of leading order and higher order contributions fit to the data and the amount
%estimated from {\footnotesize\rm ISAJET\normalsize}.
%

The cross section $d\sigma_{b\overline{b}}^{\mu\mu}/d\Delta\phi^{\mu\mu}$
is shown in Fig.~3.
Also shown are the LO and NLO QCD predictions which are determined using {\footnotesize\rm HVQJET\normalsize}
and include all subprocesses.
The grey band around the NLO prediction 
shows the combined  statistical and systematic errors associated with the 
prediction, which is $^{+74}_{-50}$\% as detailed above.
The data again show an excess above the NLO QCD
prediction but agree with the overall shape.
The agreement in shape is consistent with the presence of NLO subprocesses since the
LO prediction, which contains the smearing from the $b\rightarrow B\rightarrow\mu$
fragmentation and decay chain, does not describe the data.
%
%

%\section*{Conclusions}

In conclusion, we have measured the $b$ quark production cross section and
the $b\overline{b}$ azimuthal angle correlations using dimuons to tag the
presence of $b$ quarks. These measurements, as well as the revised inclusive
single muon measurement, are found to agree in shape with the
NLO QCD calculation of heavy flavor production but lie above the central values
of these predictions.
%
%\section*{Acknowledgments}

%%%\input{ACKNOWLEDGEMENT_PARAGRAPH.TEX}
% Acknowledgement_paragraph.tex
%
We thank the staffs at Fermilab and at collaborating institutions 
for contributions to this work, and acknowledge support from the 
Department of Energy and National Science Foundation (USA),  
Commissariat  \` a L'Energie Atomique and
CNRS/Institut National de Physique Nucl\'eaire et 
de Physique des Particules (France), 
Ministry for Science and Technology and Ministry for Atomic 
   Energy (Russia),
CAPES and CNPq (Brazil),
Departments of Atomic Energy and Science and Education (India),
Colciencias (Colombia),
CONACyT (Mexico),
Ministry of Education and KOSEF (Korea),
CONICET and UBACyT (Argentina),
A.P. Sloan Foundation,
and the Humboldt Foundation.
%
% end ack paragraph

%%%%%%%%%%%%%%%%%%%%%%%%%%%%%%%%%%%%%%%%%%%%%%%%%%%%%%%%%%%%%%%%%%%%%%%%%%%%%%%%
\newpage
{\centerline {FIGURE CAPTIONS}}
\vskip 0.5in
\noindent
{\bf Figure 1} The results of the maximum likelihood fit to the data for
(a) $p_T^{\rm rel}$ of the leading muon and (b) $r_z$.
Also included are the curves showing the contribution from each process to the
dimuon sample.

\vskip 0.5in
\noindent
{\bf Figure 2} (a) The unfolded leading muon $p_T$ spectrum for $b\overline{b}$ production compared
to the predicted spectrum (see
text) where the data errors are statistical (inner) and total (outer) and the Monte Carlo 
errors are total (shaded band); (b) the $b$ quark production cross section for $|y^b|<1.0$
compared with
the revised inclusive single muon results and the NLO QCD prediction. The error bars on the data
represent the total error.  The theoretical uncertainty shows the uncertainty
associated with the factorization and renormalization scales and the $b$ quark mass.
Also shown are the inclusive single muon data from CDF\cite{cdfb1}.

\vskip 0.5in
\noindent
{\bf Figure 3} The $\Delta\phi^{\mu\mu}$ spectrum for $b\overline{b}$ production compared to the
predicted spectrum (see text). The errors on the data are statistical and total.
The solid histogram shows the NLO prediction with the grey band
indicating the total uncertainty.
Also shown is the LO prediction (dotted histogram) with the statistical error only.

\newpage

%%%%%%%%FIG1
\begin{figure}

\vskip 0.2in
{\centerline {\bf Figure 1}}
\vskip 0.5in
%
%\vbox to 8cm{
%\vbox to 6cm{
%\vfill
%\special{psfile=bit11_ptr1_ptr2_fit_prl_new.ps
%          angle=0 hscale=40 vscale=40 hoffset=10 voffset=-160}
%\special{psfile=et5_big6.eps
%	angle=0 hscale=32 vscale=32 hoffset=35 voffset=0}
%\vfill
\centerline{\psfig{figure=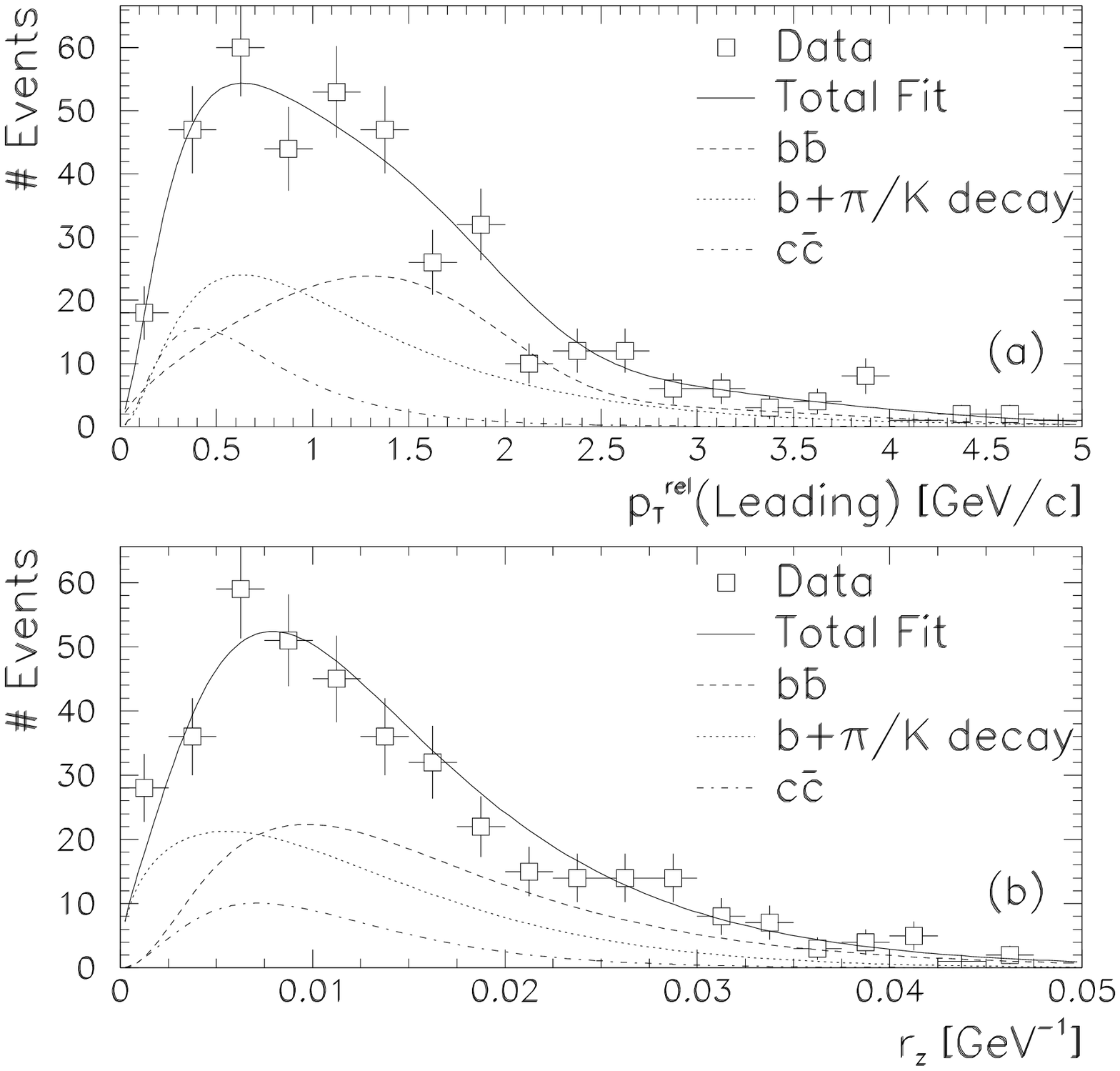,width=15cm}}
%\caption{The results of the maximum likelihood fit to the data for
%(a) $p_T^{\rm rel}$ of the leading muon and (b) $r_z$.
%Also included are the curves showing the contribution from each process to the
%dimuon sample.}
\label{fig:fit}
\vspace{0.1cm}
\end{figure}
%%%%%%%%%%%%%%%%%%%%%%

\newpage
\vskip 0.5in
{\centerline {\bf Figure 2}}
\vskip 0.5in

% %%%%%%%%FIG2
\begin{figure}                
%\vbox to 11.8cm{
%\vbox to 11.8cm{
%\vfill
%\special{psfile=fig3a.ps
%          angle=0 hscale=32 vscale=32 hoffset=25 voffset=-70}
%\special{psfile=d0_data_prl.ps
%          angle=0 hscale=32 vscale=32 hoffset=25 voffset=-215}
%\special{psfile=et1_big7.eps
%	angle=0 hscale=32 vscale=32 hoffset=35 voffset=0}
%\special{psfile=et3_big12.eps
%	angle=0 hscale=32 vscale=32 hoffset=35 voffset=-165}
%\vfill
%}
 \centerline{\psfig{figure=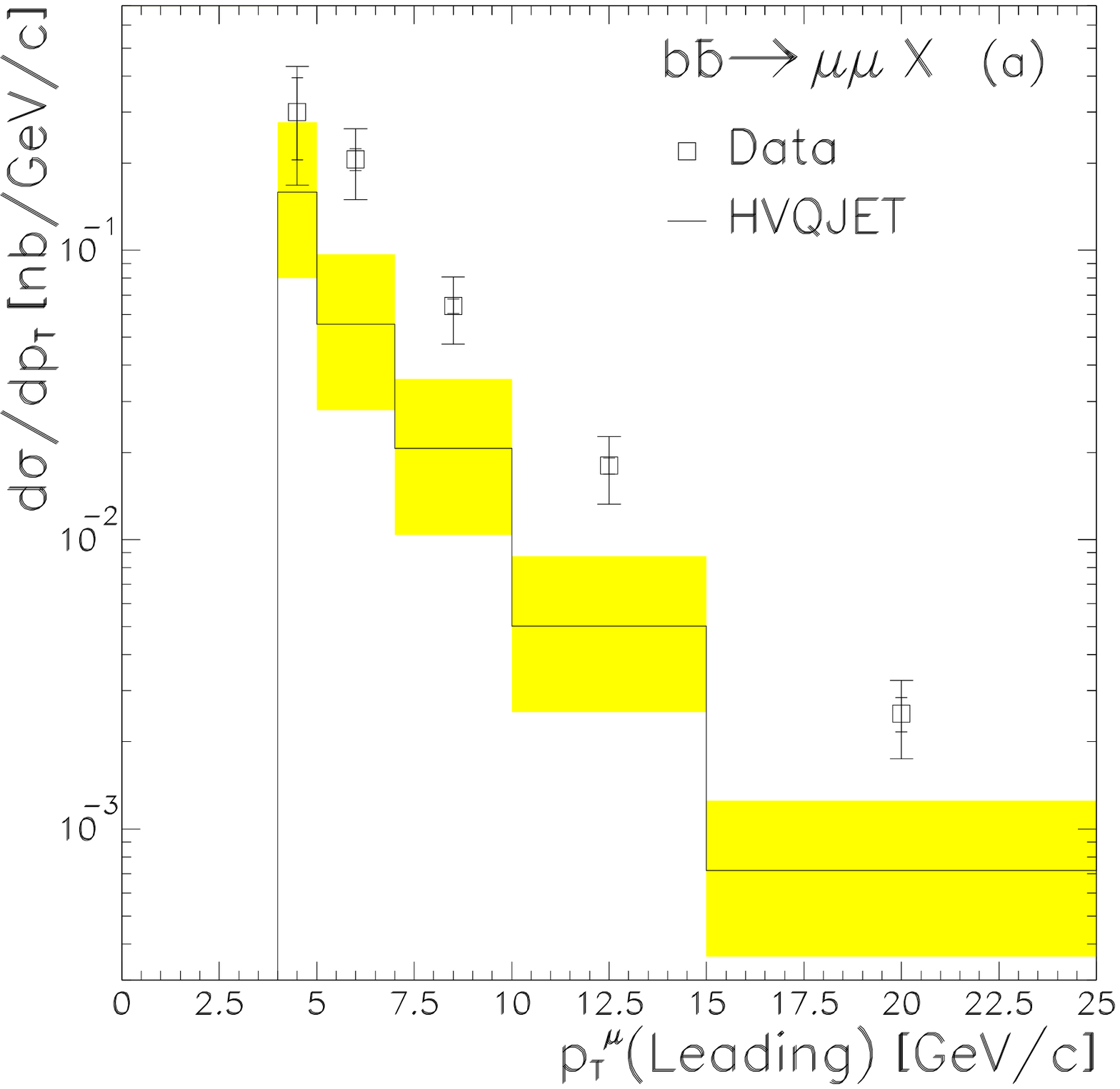,width=10cm,height=10cm}}

%  \vspace*{-9cm}

%  \hspace*{8.5cm}
 \centerline{\psfig{figure=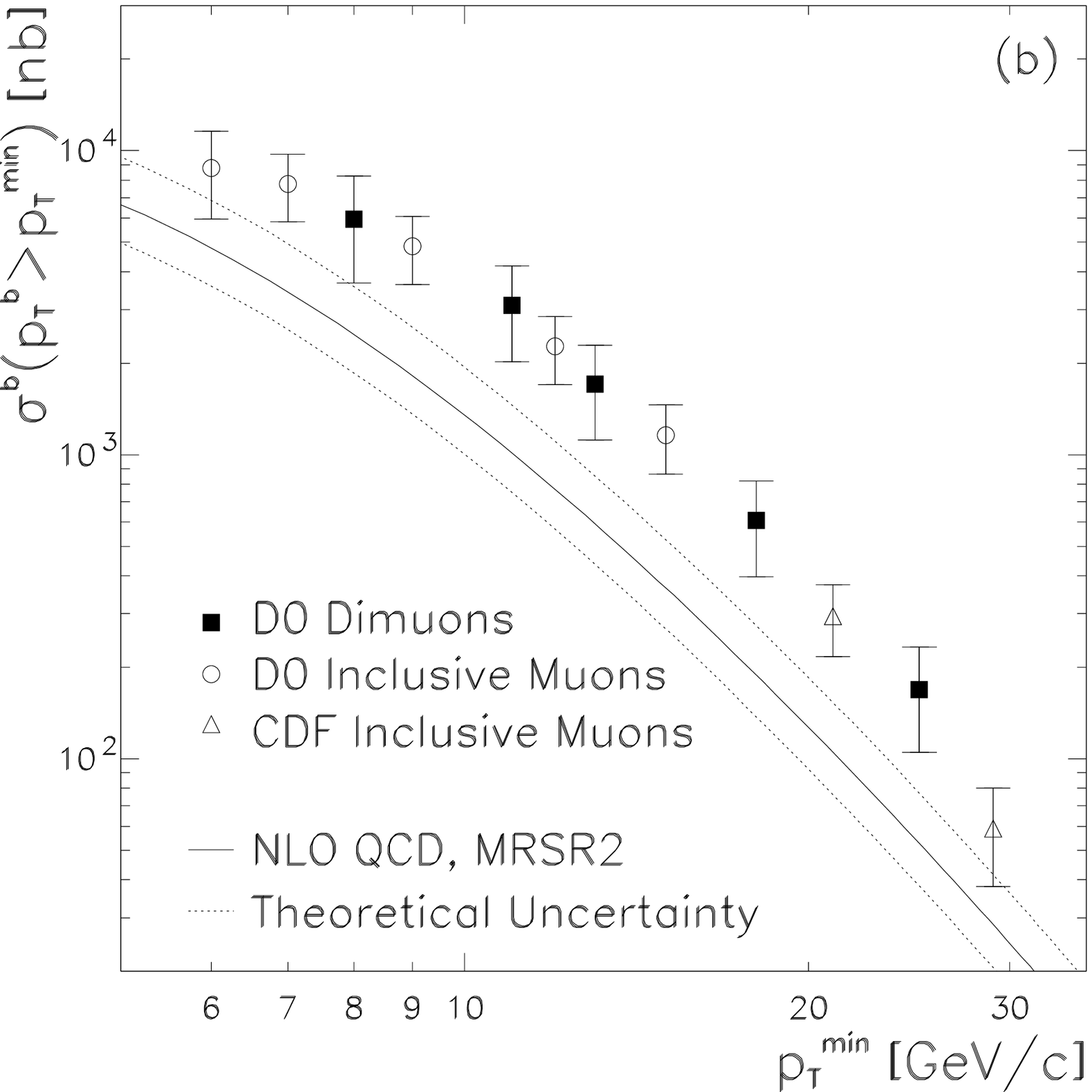,width=10cm,height=10cm}}
%\caption{(a) The unfolded leading muon $p_T$ spectrum for $b\overline{b}$ production compared
%to the predicted spectrum (see
%text) where the data errors are statistical (inner) and total (outer) and the Monte Carlo 
%errors are total (shaded band); (b) the $b$ quark production cross section for $|y^b|<1.0$
%compared with
%the revised inclusive single muon results and the NLO QCD prediction. The error bars on the data
%represent the total error.  The theoretical uncertainty shows the uncertainty
%associated with the factorization and renormalization scales and the $b$ quark mass.
%Also shown are the inclusive single muon data from CDF\cite{cdfb1}.}
\label{fig:sigma}
\vspace{0.1cm}
\end{figure}
%%%%%%%%%%%%%%%%%%%%%% %
\newpage
\vskip 0.5in
{\centerline {\bf Figure 3}}
\vskip 0.5in

%%%%%%%%FIG3
\begin{figure}
%\vbox to 9.0cm{
%\vbox to 6.0cm{
%\vfill
%\special{psfile=et2_big7.eps
%	angle=0 hscale=32 vscale=32 hoffset=35 voffset=-85}
%\special{psfile=et2_big2.eps
%	angle=0 hscale=32 vscale=32 hoffset=35 voffset=-40}
%\special{psfile=et4_big2.eps
%	angle=0 hscale=32 vscale=32 hoffset=35 voffset=-205}
%\vfill
%}
\centerline{\psfig{figure=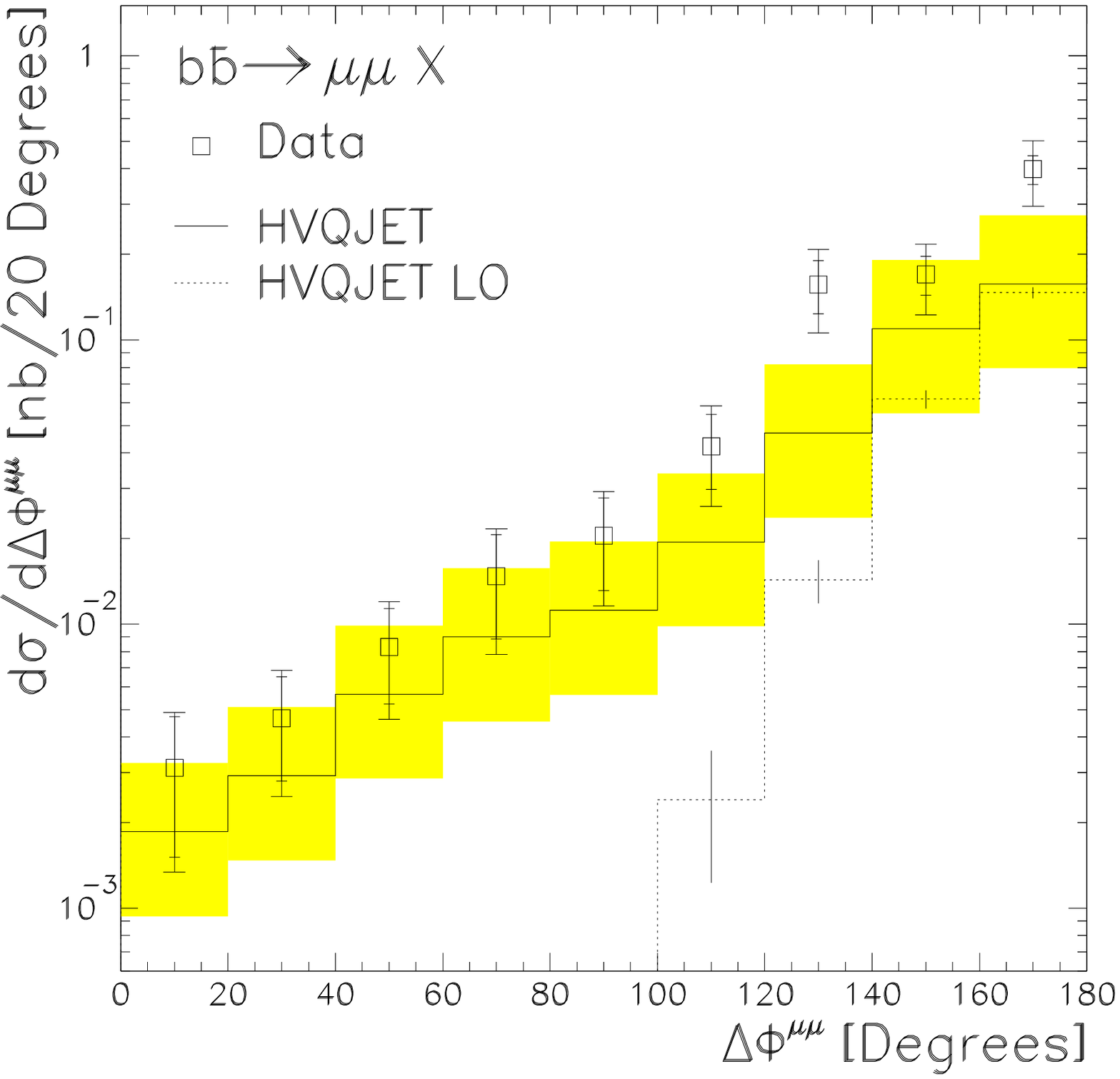,width=15cm}}
%\caption{The $\Delta\phi^{\mu\mu}$ spectrum for $b\overline{b}$ production compared to the
%predicted spectrum (see text). The errors on the data are statistical and total.
%The solid histogram shows the NLO prediction with the grey band
%indicating the total uncertainty.
%Also shown is the LO prediction (dotted histogram) with the statistical error only.}
\label{fig:mnrdphi}
\vspace{0.1cm}
\end{figure}
%%%%%%%%%%%%%%%%%%%%%%

%%%%%%%%%%%%%%%%%%%%%%%%%%%%%%%%%%%%%%%%%%%%%%%%%%%%%%%%%%%%%%%%%%%%%%%%%%%%%%%%
\newpage
{\centerline {TABLE CAPTIONS}}
\vskip 0.5in
\noindent
{\bf Table 1} Cross sections for $b\overline{b}\rightarrow\mu\mu$ production.

\vskip 0.5in
\noindent
{\bf Table 2} Results for the $b$ quark production cross section
for $|y^b|~<~1.0$.

\newpage

%% Table 1
\begin{table}
\vskip 0.2in
{\centerline {\bf Table 1}}
\vskip 0.5in
\begin{center}
\label{tab:mumu}
%\caption{Cross sections for $b\overline{b}\rightarrow\mu\mu$ production.}
\begin{tabular}{cccc}
 {$p_T^{\mu_1}$}  & {$d\sigma^{\mu\mu}/dp_T$} & {Stat Error}  & {Syst Error}\\
 {[GeV/$c$]}      & [nb/GeV/$c$]             & [nb/GeV/$c$]   & [nb/GeV/$c$]      \\ \hline
 4-5            & 0.30                     & 0.095          & 0.092 \\
 5-7            & 0.21                     & 0.018          & 0.054 \\
 7-10           & 0.064                    & 0.0037         & 0.016 \\
 10-15          & 0.018                    & 0.0012         & 0.0046 \\
 15-25          & 0.0025                   & 0.00034        & 0.00067 \\
\end{tabular}
\end{center}
\end{table}
%

%% Table 2
\newpage
\vskip 0.5in
{\centerline {\bf Table 2}}
\vskip 0.5in

\begin{table}
\begin{center}
\label{tab:bb}
%\caption{Results for the $b$ quark production cross section
%for $|y^b|~<~1.0$.}
\begin{tabular}{cccc}
 {$p_T^b$}    & {$\sigma^{b}$} & {Total Error}  & {Data Source}\\
 {[GeV/$c$]}  & [$\mu$b]       & [$\mu$b]   & \\ \hline
 6  & 8.76    & 2.80           & Single Muon \\
 7  & 7.78    & 1.95           & Single Muon \\
 8  & 5.96    & 2.29           & Dimuon \\
 9  & 4.85    & 1.22           & Single Muon \\
 11 & 3.10    & 1.08           & Dimuon \\
 12 & 2.28    & 0.58           & Single Muon \\
 13 & 1.71    & 0.59           & Dimuon \\
 15 & 1.16    & 0.30           & Single Muon \\
 18 & 0.61    & 0.21           & Dimuon \\
 25 & 0.17    & 0.06           & Dimuon \\
\end{tabular}
\end{center}
\end{table}           


\begin{references}
\bibitem{isajet2}
Particle decays based on the CLEO decay tables are implemented in
{\footnotesize\rm ISAJET\normalsize} starting with V7.22.
\end{references}
\end{document}